\documentclass[aps,prb,twocolumn,floatfix,showpacs]{revtex4-1}

\usepackage{graphicx}
\usepackage{amsmath,amssymb}
\usepackage{dcolumn} 
\usepackage{bm} 
\usepackage{xcolor}
\usepackage[utf8]{inputenc}

\usepackage{hyperref}
\hypersetup{%
  colorlinks=true,
  citecolor=blue,
  urlcolor=blue,
  linkcolor=blue,
  }

\begin{document}

\title{Three-particle states and brightening of intervalley excitons in a doped MoS$_2$ monolayer}

\author{Y.V. Zhumagulov$^{1,2}$}
\author{A. Vagov$^{3,1}$}
\author{N.Yu. Senkevich$^{1}$}
\author{D.R. Gulevich$^{1}$}
\author{V. Perebeinos$^{4}$}


\affiliation{$^{1}$ITMO University, St. Petersburg 197101, Russia}
\affiliation{$^{2}$University of Regensburg, Regensburg, 93040, Germany}
\affiliation{$^{3}$Institute for Theoretical Physics III, University of Bayreuth, Bayreuth 95440, Germany}
\affiliation{$^{4}$Department of Electrical Engineering, University at Buffalo, The State University of New York, Buffalo, NY 14260, USA}

\email{vasilipe@buffalo.edu}

\date{\today}

\begin{abstract}
Optical spectra of two-dimensional transition-metal dichalcogenides (TMDC) are influenced by complex multi-particle excitonic states. Their theoretical analysis requires solving the many-body problem, which in most cases, is prohibitively complicated. In this work, we calculate the optical spectra by exact diagonalization of the three-particle Hamiltonian within the  Tamm-Dancoff approximation where the doping effects are accounted for via the Pauli blocking mechanism, modelled by a discretized mesh in the momentum space. The single-particle basis is extracted from the {\it ab initio} calculations.  Obtained three-particle eigenstates and the corresponding transition dipole matrix elements are used to calculate the linear absorption spectra as a function of the doping level. Results for negatively doped MoS$_2$ monolayer (ML) are in an excellent quantitative agreement with the available experimental data, validating our approach. The results predict additional spectral features due to the intervalley exciton that is optically dark in an undoped ML but is brightened by the doping. Our approach can be applied to a plethora of other atomically thin semiconductors, where the doping induced brightening of the many-particle states is also anticipated. 
\end{abstract}

\maketitle

\section{Introduction}

MoS$_2$ ML's are non-centrosymmetric 2D semiconductors with two degenerate direct gaps in the single-particle spectrum at the $\pm K$ points of the Brillouin zone.~\cite{Mak2010, Ramasubramaniam2012, Xiao2012, Kormnyos2015} Such structures have many properties that are of interest to both fundamental research and practical applications. In particular, they are strongly coupled to light, yielding strong photoluminescence,~\cite{Mak2010, Splendiani2010} and have a large spin-orbit interaction, which allows one for efficient manipulation of their spin and valley degrees of freedom.~\cite{Mak2012nano, Zeng2012, Cao2012, Sallen2012} A unique combination of optical and electrical characteristics of those materials makes them very attractive for a variety of optoelectronic applications~\cite{Xia2014,Wang2018} including logic circuits,~\cite{Radisavljevic2011,Wang2012} phototransistors,~\cite{LopezSanchez2013} light sensors~\cite{Perkins2013} as well as light-producing and harvesting devices.\cite{Feng2012,Cheng2014,LopezSanchez2014,Pospischil2014,Ross2014}

The 2D geometry of such materials enhances the Coulomb interaction,  giving rise, in particular,  to a much larger exciton binding energy~\cite{Mak2010, Splendiani2010, Komsa2012, Feng2012, Qiu2013, BenAmara2016} in comparison with that in bulk semiconductors.~\cite{Yu2010} This enhancement also facilitates other many-body states including three-particle charged excitons or trions.~\cite{Mak2012, Ross2014, Lui2014, Rezk2016, Zhang2014, Mouri2013, Singh2016, Scheuschner2014, Soklaski2014, Zhang2015_2} Signatures of negatively charged trions, consisting of two electrons and one hole, were previously observed in the photoluminescence spectrum of a field-effect transistor made of a MoS$_2$ ML.~\cite{Mak2012}  It must be noted that the existence of two degenerate valleys in the band structure results in many different types of exciton and trion states, such that a reliable interpretation of experiments is possible only alongside comprehensive theoretical analysis of three particle states.~\cite{Wang2018}

A direct solution of a three-body problem requires enormous computation efforts even for a 2D system and, especially, in the presence of other particles due to the doping. These difficulties forced researchers to look for approximate methods, such as the variational~\cite{Berkelbach2013,Kezerashvili2016} and stochastic approaches,~\cite{Kidd2016,Zhang2015,VanTuan2018}  perturbation expansions,~\cite{Efimkin2017,Sidler2016,Back2017,Scharf_2019} path integral~\cite{Kylnp2015,Velizhanin2015} and diffusion Monte Carlo methods.~\cite{Mayers2015} Recently, a direct solution of the three-body problem within the Tamm-Dancoff approach become possible.~\cite{Deilmann2016,Drppel2017,Deilmann2017,Deilmann2018,Deilmann2018BP,Arora2019,Torche2019,Tempelaar2019}
However, solving the three-body problem itself is of limited use, if one wants to describe real experiments, where trions are excited in a doped structure where they interact with the excess carriers.~\cite{Mak2012, Ross2013, Mouri2013, Smoleski2019}


In this work, we overcome this shortcoming by calculating the trion states from the exact solution of the three-particle Hamiltonian, obtained within the Tamm-Dancoff approximation, where the discretized mesh in the momentum space is introduced to account for the Pauli blocking due to the doping. The set of single particle basis states is obtained from the {\it ab initio} calculations of the electronic band structure. The solution to the three-particle problem gives the energy spectrum and wavefunctions of the trion states used to calculate transition dipole matrix elements and then the linear absorption spectra as a function of doping level. Results for the negatively charged trion states in an electrically doped  MoS$_2$ ML are in excellent quantitative agreement with the available experimental data. Our calculations predict that the intervalley exciton state, which is dark in undoped ML's, becomes optically active (brightens) when the ML is sufficiently doped.

The paper is organized as follows. The main steps of the calculations are explained in Sec.~\ref{sec:models}, which describes the solution of the three-body problem in the Tamm-Dancoff approximation and optical spectra calculations in the presence of doping. Results for the optical spectra as a function of doping and dielectric environment are presented in Sec.~\ref{sec:results}. The relation of the low energy trion states with the corresponding two-particle exciton states in undoped ML's is also discussed in  Sec.~\ref{sec:results}. We summarize our findings in Sec.~\ref{sec:conclusions} and outline future directions.


\section{Model and Methods}
\label{sec:models}

\subsection{Three-particle Hamiltonian}
\label{sec:trions_hamiltonian}

The calculations of the three-particle states and the related optical spectra are done as follows. First, we obtain single-particle states or the band structure of a MoS$_2$ monolayer using a standard {\it ab initio} approach that combines the density functional theory (DFT) and the GW method where the spin-orbit interaction is taken into account within the first-order perturbation theory.~\cite{Olsen2016} 

Obtained single-particle states are used as a basis set for the many-body Hamiltonian, where we take into account the Coulomb interaction between electrons in the conduction and holes in the valence bands. Three-particle states are constructed  as linear combinations
\begin{align}
  \left| t \right\rangle = \sum_{c_1,c_2,v} A_{c_1 c_2 v}^{t}   a_{c_1}^\dagger  a_{c_2}^\dagger a_{v}^\dagger \left| 0 \right\rangle,
  \label{eq:expansion}
\end{align}
where the index $c_{1,2}/v$ denote electron/hole states in the conduction/valence bands and the double counting is avoided by imposing the restriction $c_1<c_2$. The corresponding three-particle wavefunction is obtained from the single-particle ones $\phi_{c,v}(x)$ as
\begin{align}
    \Psi^{t}(&x_1,x_2,x_3) = \frac{1}{\sqrt{2}} \sum_{c_1,c_2,v} A_{c_1 c_2 v}^{t} \phi_{v}^{*}(x_3) \notag \\
    &\times \big[ \phi_{c_1}(x_1)\phi_{c_2}(x_2)-\phi_{c_2}(x_1)\phi_{c_1}(x_2) \big].
\label{eq:wave_function}
\end{align} 
Coefficients $A_{c_1 c_2 v}^{t}$ of the expansions (\ref{eq:expansion}) and (\ref{eq:wave_function}) are found by solving the matrix eigenvalue problem 
\begin{align}
\label{eq:eigen}
\sum_{c_1^\prime c_2^\prime v^\prime} {\cal H}_{c_1c_2v}^{c_1^\prime c_2^\prime v^\prime} A_{c_1^\prime c_2^\prime v^\prime}^{t} = \varepsilon_t A_{c_1c_2 v}^{t},
\end{align}
where the matrix Hamiltonian in the Tamm-Dancoff approximation has three contributions ${\cal H} = {\cal H}_{0} + {\cal H}_{c}  + {\cal H}_{v}$, defined as 
\begin{align}
\label{eq:Hamiltonian}
& {\cal H}_{0}=(\varepsilon_{c_1} + \varepsilon_{c_2} -\varepsilon_{v}) \delta_{c_1}^{c_1^\prime} \delta_{c_2}^{c_2^\prime}  \delta_{v}^{ v^\prime}, \notag \\
&{\cal H}_{c} =(W_{c_1c_2}^{c_1^\prime c_2'}-W_{c_1c_2}^{c_2^\prime c_1^\prime})\delta_{v}^{v^\prime}, \nonumber \\
&{\cal H}_{v}=-(W_{v'c_1}^{vc_1^\prime}-V_{v^\prime c_1}^{c_1^\prime v})\delta_{c_2}^{c_2'}-(W_{v^\prime c_2}^{vc_2^\prime}-V_{v^\prime c_2}^{c_2^\prime v})\delta_{c_1}^{c_1^\prime} \nonumber \\
& \quad \quad \, \, +(W_{v^\prime c_1}^{vc_2^\prime}-V_{v^\prime c_1}^{c_2^\prime v})\delta_{c_2}^{c_1^\prime }+(W_{v^\prime c_2}^{vc_1^\prime }-V_{v^\prime c_2}^{c_1^\prime v})\delta_{c_1}^{c_2^\prime },
\end{align}
where $\varepsilon_c $ and $\varepsilon_v$  are energies of the single-particle electron and hole states. Matrix elements for the bare Coulomb interaction $V$ are given by
\begin{equation}
\label{eq:Coulomb}
    V^{ab}_{cd} = V(\textbf{k}_a -\textbf{k}_c) \langle u_c | u_a \rangle \langle u_d | u_b \rangle,
\end{equation}
where $V ({\bf q}) = 2\pi e^2/ q$ and $\langle u_c| u_a \rangle$ is the overlap of the single-particle Bloch states $c$ and $a$ and $\varepsilon$ is the effective dielectric constant that depends on the environment. The screened potential is given also by Eq.~(\ref{eq:Coulomb}), however, instead of the bare Coulomb potential $V({\bf q})$ we use the standard Rytova-Keldysh  expression~\cite{rytova1967the8248,keldysh,Cudazzo2011}
\begin{align}
\label{eq:Coulomb_screened}
   W({\bf q}) =  \frac{2\pi e^2}{\varepsilon q(1+r_0 q)},
\end{align}
where the $r_0$ is the screening length~\cite{Wu2015}.

Finally, we  also calculate two-particle neutral exciton states in the undoped ML using the standard approach based on the Bethe-Salpeter equation (BSE) for comparison.~\cite{PhysRevB.62.4927} This yields the benchmark for the results obtained for the doped ML's in the limit of small doping.    

\begin{figure}
\includegraphics[width=0.47\textwidth]{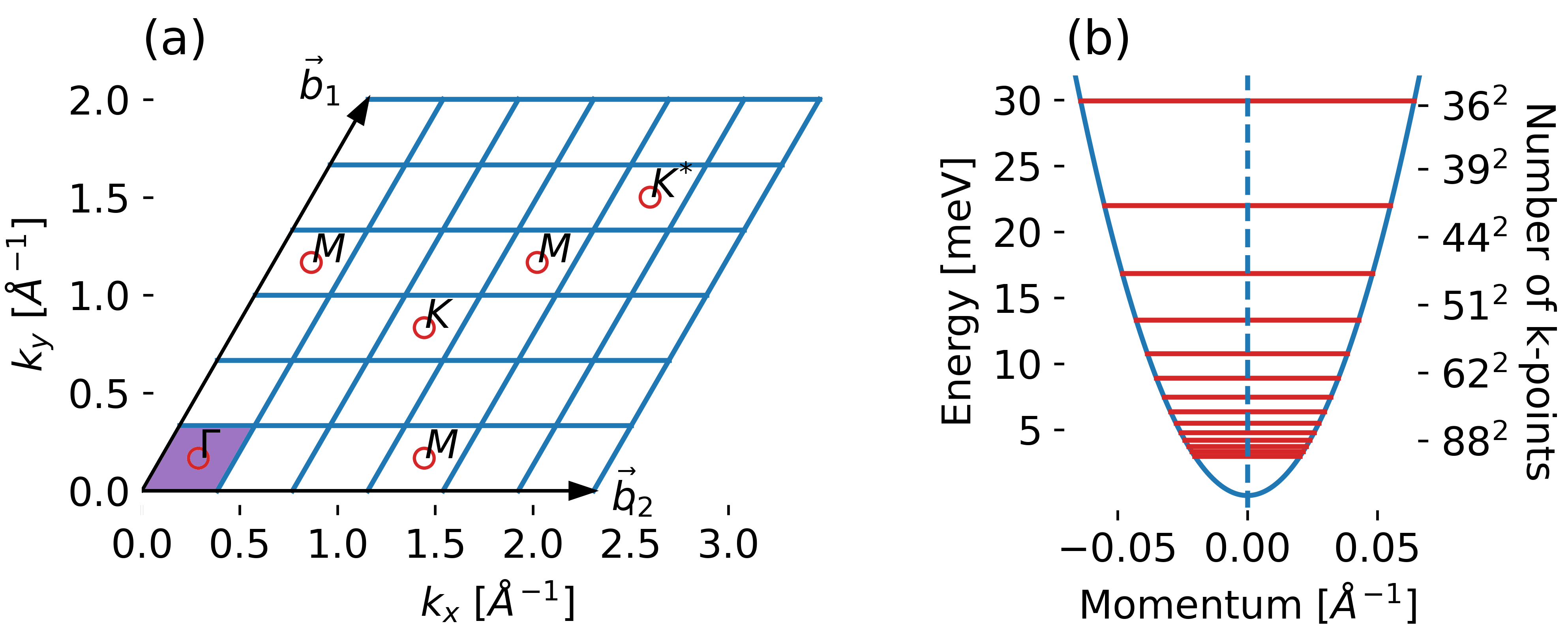}
\caption{(a) Schematic discretization of the Brillouin zone with the basis vectors $\vec{b}_{1,2}$ and the $6\times 6$ k-point mesh. A magenta filled polygon shows an elementary area $\Delta k^2\sqrt{3}/2$ per k-point. Points $\Gamma$, $K$,$K^{\prime}$ and $M$ are shown for reference. (b) Relation between the Fermi momentum (x axis), the Fermi energy of the doping electrons (left y axis) and the number of k-points of the mesh (right y axis).}
\label{fig:doping}
\end{figure}

\subsection{Doping influence}
\label{sec:doping}

Calculating three particle states cannot be used to describe optical properties of doped MoS$_2$ ML's, where excess electrons interact with the carriers composing trions. The analysis of trion states in a doped ML with excess electrons requires very non-trivial many-body calculations. However, it can be solved with an acceptable accuracy by using a numerical scheme which is no more complex than the original three-body problem. The main idea is to relate the doping with the discrete momentum space. The discretization is done in all numerical calculations, however, additional effects introduced by it are commonly regarded as an artifact that must be eliminated by choosing a sufficiently small discretization interval $\Delta k$. At the same time, the discretization is directly related to the doping density which can be intuitively understood by recalling that the discretization in the ${ k}$-space is equivalent to considering a system in a finite box of size $L$. The discretization can thus be loosely interpreted as each $L$-sized box in the periodic system has a trion and, hence, a single excess electron. For the $N\times N$ k-point mesh of the Brillouin zone, shown in Fig.~\ref{fig:doping}a,  this yields an estimate for the doping density $n = g_s g_v/(AN^2)$, where  $g_s=g_v=2$ are spin and valley degeneracies, respectively, $A=\sqrt{3}a^2/2$ is the unit cell area, and $a$ is the lattice constant.

One can also see the relation between the discretization and the doping from a different perspective. The excess electrons are Fermi particles, which, following the Pauli's exclusion principle, occupy all the single-particle states below the Fermi level $E_F$ (if we assume that the system temperature $k_BT << E_F$). The interaction-induced scattering to the occupied states is thus blocked and such states must be excluded from the solution of the problem. The discretization in the $k$-space effectively introduces the Pauli-blocking by restricting the available phase-space to states with energy $E\ge E_F=\hbar^2 \Delta k^2/2 m_c^\ast$, where $\Delta k=4\pi/(\sqrt{3}aN)$ is related to the area of the Brillouin zone $(N\Delta k)^2\sqrt{3}/2=4\pi^2/A$, see Fig.~\ref{fig:doping}b. In this estimate we use  effective mass of the conduction band $m_c=0.52 m_e$, which is consistent with the earlier works.~\cite{Xiao2012,Berkelbach2013}
Note, that the usual relationship $E_F=\hbar^2\pi n/2m_c$ between the Fermi energy and the carrier density does not hold in our case, because an effective Fermi area on a discretized mesh is $\Delta k^2\sqrt{3}/2$ and not $\pi \Delta k^2$.

\begin{figure}[t!]
\begin{center}
\includegraphics[width=\linewidth]{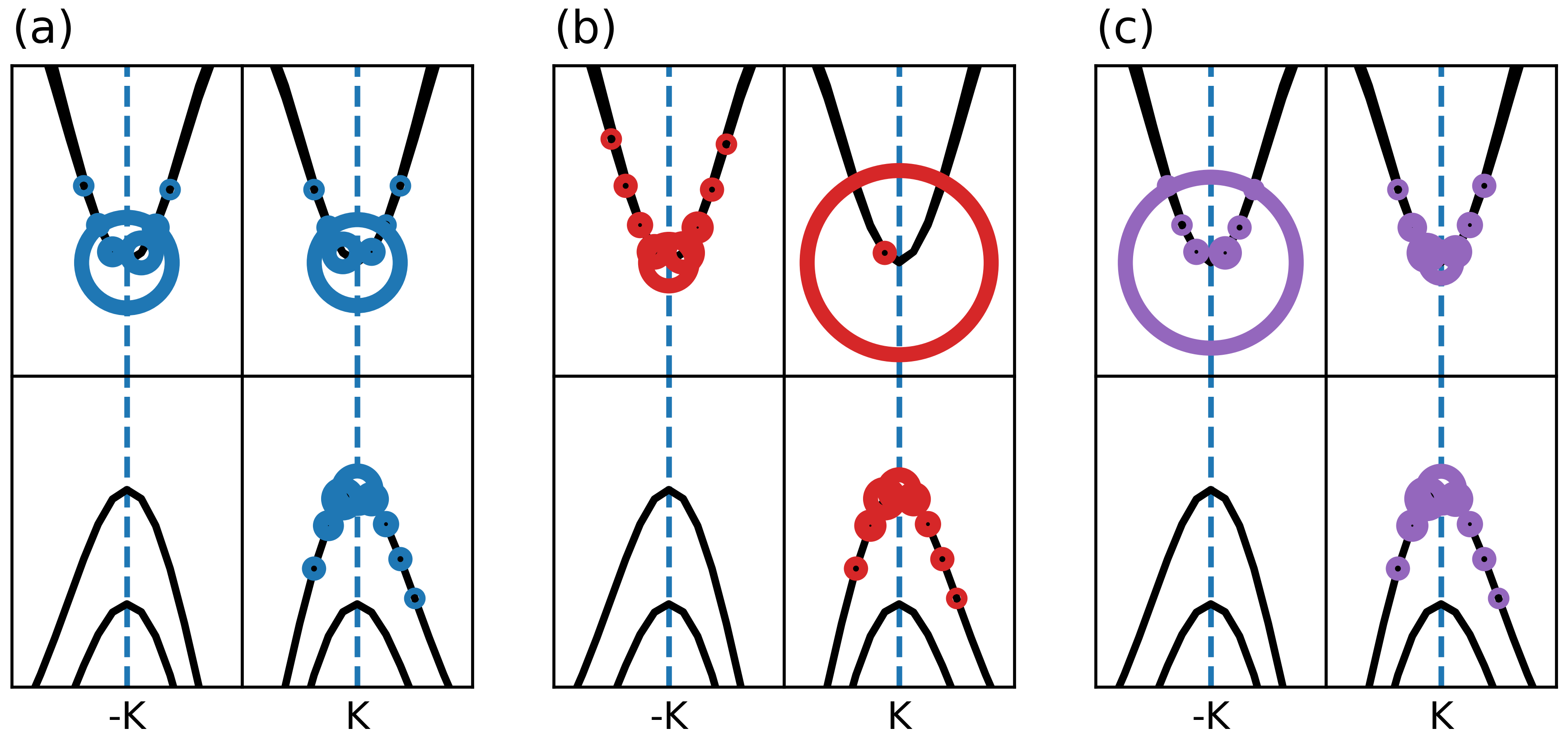}
\end{center}
\caption{ Band structure of a MoS$_2$ ML near points $K$ and $-K$ of the Brillouin zone. The single particle density of states (DOS) for trion wave functions is illustrated by circles: their centres point to the contributing states and the radii give the contribution weights. Panels a), b) and c) correspond to different three-particle states $X^-$, $iX^0_e$ and $X^0_e$, schematically illustrated, where longer orbitals depict weakly bound electrons. Three-particle states $X^0_e$ and $iX^0_e$ are related, respectively, to the intervalley $X^0$ and intravalley $iX^0$ two-particle excitons in the undoped ML. The calculations are shown for the Fermi level $E_F=3.04$ meV. }
\label{fig:band_structure}
\end{figure}

\subsection{Absorption spectrum}
 \label{sec:absorption}

The linear absorption spectrum is calculated as a sum of the transition rates between all possible free electron and three-particle states, which yields the expression:
\begin{align}
    L(\varepsilon ) \propto \sum_{c,t} 
    \big | \langle t | \hat {\bf P} | c  \rangle \big|^2 \delta \left( \varepsilon - \varepsilon_t + \varepsilon_c \right), 
    \label{eq:spectrum}
\end{align}
where the summation runs over single-electron  $c$ and three-particle $t$ states with the same momenta $\bf k$.  The transition matrix element in this expression is calculated as 
\begin{align}
    &\left\langle t \right| \hat {\bf P} \left| c \right \rangle = \sum_{c_1c_2 v}A^{t}_{c_1c_2v} \left( \textbf{p}_{c_1v}\delta_{c}^{c_2} - \textbf{p}_{c_2v}\delta_{c}^{c_1}\right),
\label{eq:cutoff}
\end{align}
and ${\bf p}_{cv}$ denotes the dipole matrix element calculated for conduction  $c$ and valence $v$ states. In order to account for the finite life-time of three-particle states the delta function in Eq.~(\ref{eq:spectrum}) is replaced by the Gaussian function of the finite width, here we assume this width $1$ meV.

\section{Results}
\label{sec:results}

\subsection{Band structure and effective tight-banding Hamiltonian}
\label{sec:band_structure}

The band structure of a single-layered MoS$_2$ is calculated using the DFT approach as implemented in the GPAW~\cite{HjorthLarsen2017,Enkovaara2010} code with the PBE exchange-correlation functional.~\cite{Perdew1996} The spin-orbit interaction is taken into account within the first-order perturbation contribution.~\cite{Olsen2016} The calculations are done using the plane-wave basis with the $12 \times 12 \times 1$ grid in the $k$ space with the energy cutoff $600$ eV. The lattice constant of MoS$_2$ is assumed $a=3.14$ \AA, the vacuum distance between MoS$_2$ layers is $20$ \AA. We also assume that the MoS$_2$ ML is placed on the  SiO$_2$ substrate, the effective dielectric constant of the ML on this substrate is $\varepsilon =2.45$.~\cite{Wu2015,Cudazzo2011,Berkelbach2013}

Obtained band structure is similar to the one calculated in earlier works,~\cite{Xiao2012,Qiu2013,Drppel2017} it has two degenerate direct gaps $E_g = 2.184$ eV at two valley points $K$ and $-K$ of the Brillouin zone.
Figure~\ref{fig:band_structure} shows the band structure of the single-particle in the vicinity of these two points that contribute most to the lowest energy three-particle states. 

As in Ref.~\onlinecite{Scharf2016} we use the obtained band structure to construct an effective tight-binding Hamiltonian of the system. This is done by employing a standard algorithm, implemented in the Wannier90 code~\cite{Mostofi2014} with the $24 \times 24 \times 1$ mesh of the Brillouin zone. The resulting Hamiltonian has 22 bands, of which 10 conduction bands correspond to Mo atoms and 6 valence bands correspond to each of the S atoms in the elementary cell. We also apply a scissor procedure with $\Delta_{scissor}=0.497$ eV  to get the correct energy of the ground exciton state. Eigenstates of the tight-binding Hamiltonian are then used as a basis set of single-particle states in the analysis of trions.

\subsection{Three-particle states}
\label{sec:trion_states}

In the calculations of the three-particle states we assume the screening length of the Keldysh-Rytova potential to be $r_0 = 33.875$\AA$/ \varepsilon$.~\cite{Wu2015} The computational load of solving the three-body problem is reduced considerably when  the single-particle basis is restricted by constraints
\begin{align}
    \big\{ |{\bf k}_{c_1}\pm{\bf k}|, |{\bf k}_{c_2}\pm {\bf k}|, |{\bf k}_{v}\pm{\bf k}| \big\} \le q_{c},
\end{align}
where $\bf k$ is the momentum of a three-particle state and the cutoff $q_{c} = 0.4$\AA$^{-1}$ is chosen to ensure the optimal accuracy and convergence of the calculations.

The calculations reveal three qualitatively different lowest energy three-particle states. Their structure is illustrated in Fig.~\ref{fig:band_structure}, which demonstrates a contribution of the single-particle state to the three-particle wavefunction, i.e. single particle density of states (DOS), by a circle with the centre pointing to the contributing single-particle state and the radius proportional to the contribution weight. 

\begin{figure*}[t!]
\begin{center}
\includegraphics[width=\textwidth]{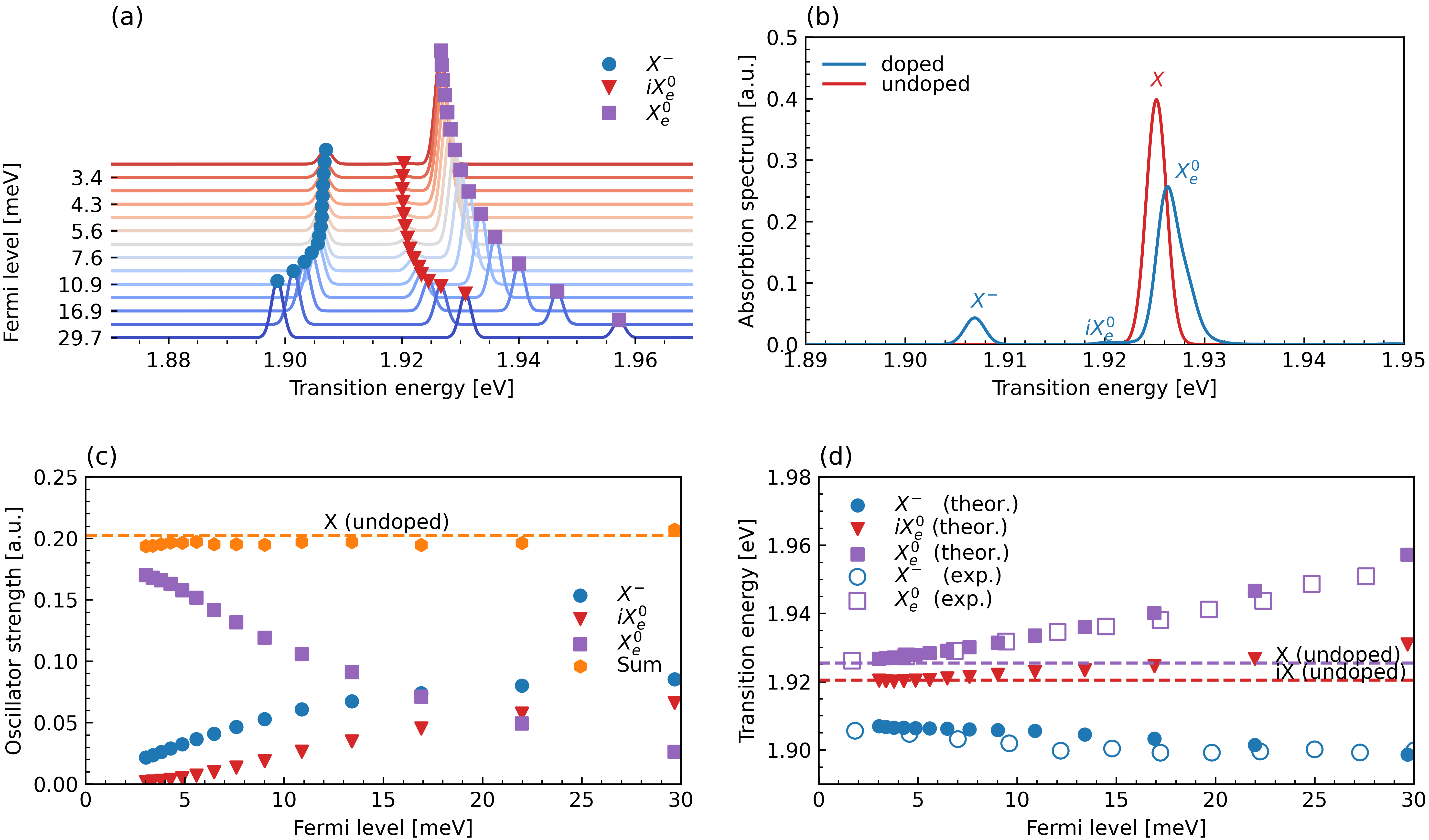}
\end{center}
\caption{(a) Absorption spectrum calculated for different values of the doping density $n$ (right) corresponding to the Fermi levels $E_F$ (left), the lines are offset vertically for clarity. The three peaks correspond to three-particle states $X^{-}$, $iX^0_e$ and $X^0_e$ shown in Fig.~\ref{fig:band_structure}. (b) Absorption spectra of the  undoped (red) and doped (blue) ML's at the Fermi level $E_F=3.04$ meV. (c) The oscillator strength (OS) of the spectral peaks and the total OS (the sum for the three peaks) as a function of the Fermi level.  Dashed line gives the OS of the $X$ exciton peak of the undoped ML. (d) The energy-position  of the peaks as a function of the Fermi level, experimental data~\cite{Mak2012} for the two peaks energies $X^-$ and $X^0_e$ are shown for comparison. Dashed lines show the intervalley $iX^0$ and intravalley $X^0$ exciton energies from the BSE solution in the undoped ML.}
\label{fig:spectra}
\end{figure*}

An example of the structure of the lowest energy trion state $X^-$ is shown in Fig.~\ref{fig:band_structure}a. There are several (four) such trion states which differ by the spin and valley of the contributing electronic states. When the doping level is small, these states can be classified as intervalley and intravalley trions.~\cite{Jones2015,Plechinger2016,Lyons2019} Three of them are optically active (bright) and can contribute to the optical spectra.~\cite{Zhumagulov2020b} In MoS$_2$ ML their energies are very close (the energy separation are of the order of $1$ meV) and, consequently, they cannot be distinguished in the spectrum unless the magnetic field is applied. In this work we do not discuss differences in their structure in detail. 

One feature is, however, shared by all these lower energy $X^-$ trion states. Figure~\ref{fig:band_structure}a shows that they comprises many electronic states of similar weights pointing to a strong localization of both electrons. In contrast, the DOS of the higher energy three-particle states, $iX^0_e$ and $X^0_e$  in Fig.~\ref{fig:band_structure}b and c, is  non-symmetric and singular: the contribution of one of the valleys is restricted to essentially one single-particle state, which means that one of the electrons is only weakly localized.

\begin{figure}
\includegraphics[width=0.47\textwidth]{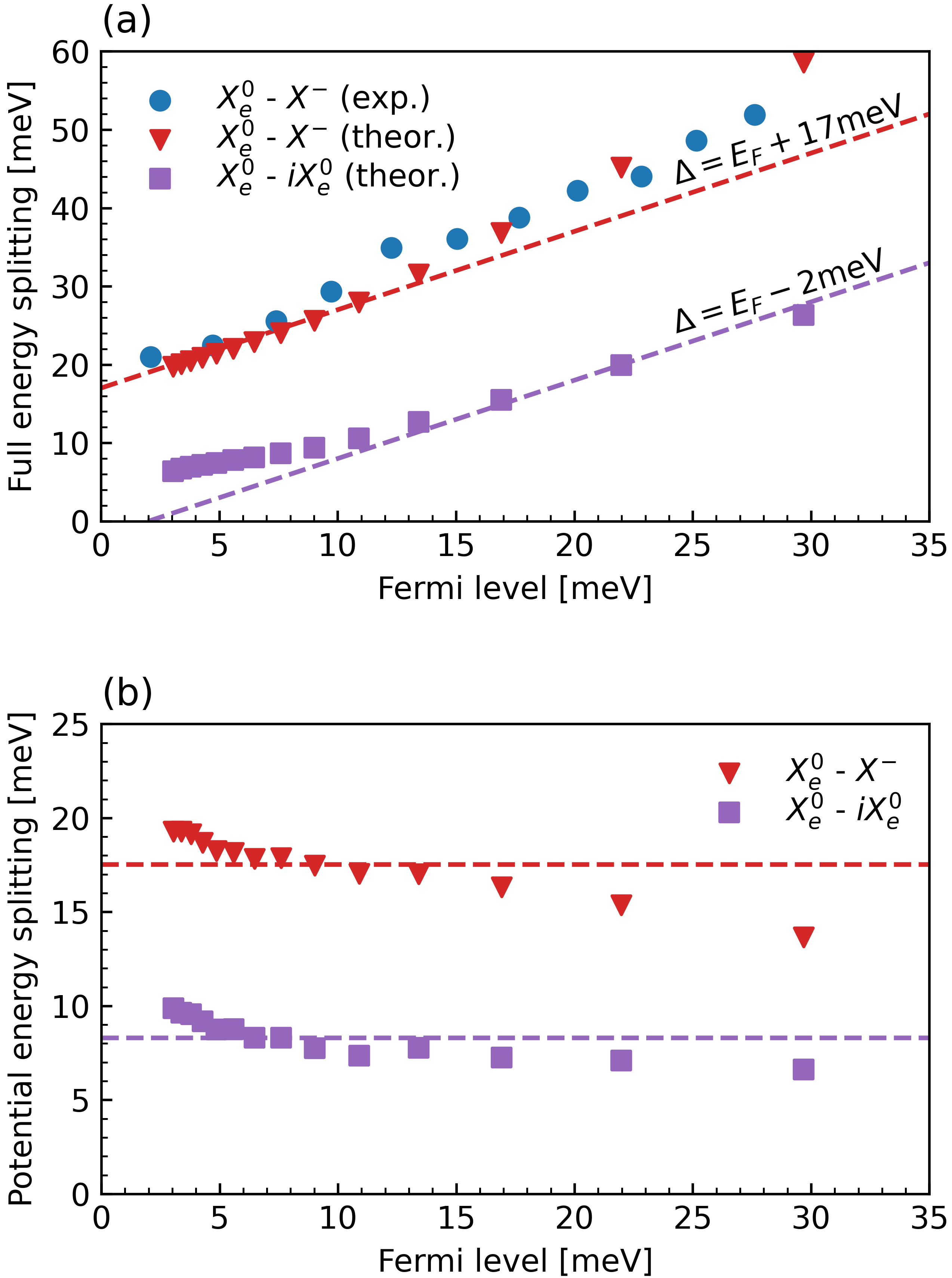}
\caption{(a) Energy differencies $\Delta E$ (splitting) for pairs of states: $X^0_e - X^{-}$ and $X^0_e - iX^0_e$, are plotted as a function of the Fermi energy $E_F$. Dashed lines illustrate the linear dependence $\Delta  E = \Delta_0 + \alpha E_F$ with $\alpha=1$. Available experimental data~\cite{Mak2012} for the energy difference of states $X^0_e - X^{-}$ are also shown for comparison. 
(b) The doping dependence of the potential energy contribution to the splitting, that yields the deviations from the linear dependence in the panel (a) (see text). The horizontal dashed lines in (b) are a  guide to the eye.}
\label{fig:separation}
\end{figure}

Three-particle states $iX^0_e$ and $X^0_e$ are, in fact, very close to the two-particle excitons, ``polarized'' by the weak coupling to the doping electrons. In principle, effects of such polarization can be described within the perturbation theory.~\cite{Efimkin2017} The BSE calculations show that two types of the corresponding bound excitonic states exist in undoped MoS$_2$ ML's: the intravalley $X^0$ exciton, where the electron and hole are both in the same valley, and the intervalley $iX^0$ exciton, where the electron and hole are in different valleys. Our solution of the three particle problem demonstrates that in the limit of vanishing doping the wave functions of  the $X^0_e$ and $iX^0_e$ states are represented as a product of the wave function of the exciton state, $X^0$ or $iX^0$, respectively, and the wave function of a delocalized electron from the other valley, with the energy being the sum of the energies of the exciton and the free electron. This relation explains the choice of notations $X^0_e$ and $iX^0_e$ which mean a neutral exciton, $X^0$ or $iX^0$, augmented by an extra  weakly bound electron. The structures of these states are illustrated schematically in Fig.~\ref{fig:band_structure}.

\begin{figure*}[t!]
\begin{center}
\includegraphics[width=\linewidth]{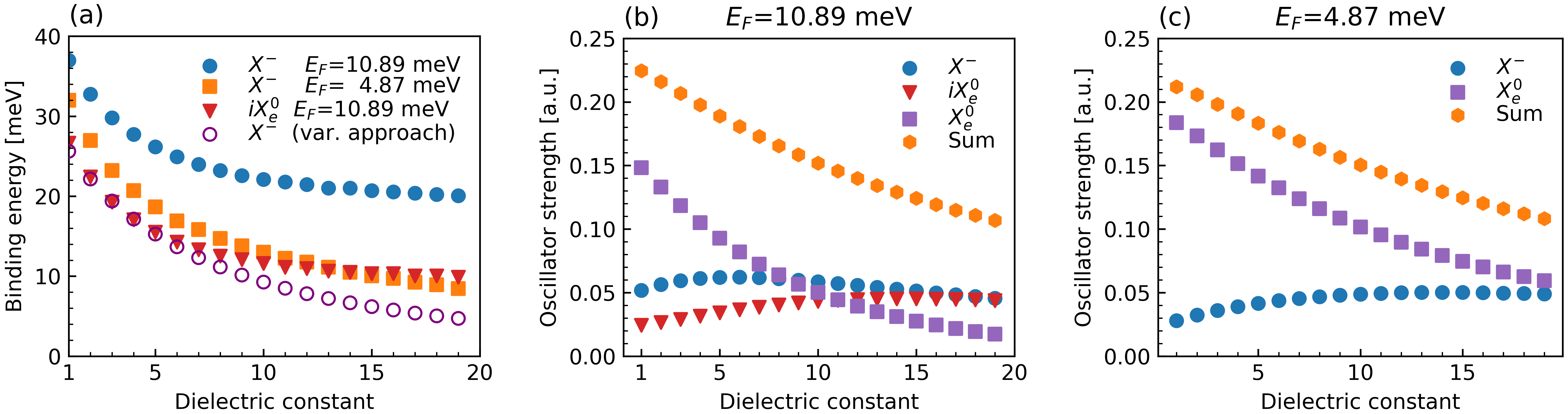}
\end{center}
\caption{ (a) Binding energy of the trion state relative to the energy of the $X^0$ exciton in the undoped ML as a function of the substrate dielectric constant $\varepsilon$, calculated at $E_F=10.89$ meV and $E_F=4.87$ meV. Results for the variational calculations (magenta circles) for the lowest $X^-$ states are given for comparison. (b) and (c)  OS for the spectral peaks, calculated at $E_F=10.89$ meV and $E_F=4.87$ meV (in the last case the $iX^0_e$ peak is not visible).}
\label{fig:epsilon}
\end{figure*}

\subsection{Absorption spectrum}
\label{sec:spectrum}

The low frequency part of the linear absorption spectrum of a MoS$_2$ ML on a SiO$_2$ substrate, calculated at several values of the doping level, is plotted in Fig.~\ref{fig:spectra}a.  The spectrum generally has three well defined peaks corresponding to  the $X^{-}$, $iX^0_e$ and $X^0_e$ states (below also referred to as the $X^{-}$, $iX^0_e$ and $X^0_e$ peaks for brevity). As noted above all lowest trion states $X^{-}$ yield a single peak in the spectrum.  

The intensity of the peaks changes with the doping level considerably. In the limit of vanishing doping the $X^0_e$ peak dominates while the peaks $X^{-}$ and $iX^0_e$ disappear. This is also illustrated in Fig.~\ref{fig:spectra}b that plots the spectrum of a doped ML with $E_F = 3.04$ meV alongside the spectrum of the undoped ML, obtained by solving the BSE for the exciton states. One sees that at small doping the $X^0_e$  peak approaches the $X^0$ exciton peak in the undoped ML.

A detailed doping dependence of the peak intensities, measured by their oscillator strength (OS), is shown in Fig.~\ref{fig:spectra}c. The results clearly demonstrate that the intensity of both $X^{-}$ and $iX^0_e$ peaks is a monotonically increasing function of $n$,  vanishing in the undoped ML limit, $n\to 0$. In contrast, the intensity of the $X^0_e$ peak is maximal at zero doping,  decreasing monotonically when $n$ rises.

Absence of the $iX^0_e$ peak at $n\to 0$ does not imply that the  state $iX^0_e$  does not exist in this limit. However, the corresponding optical transitions are suppressed, which can also be seen from the BSE calculations for the two-particle $iX^0$ excitonic state in the undoped ML, which confirms the intervalley $iX^0$ exciton being optically dark. However, when the system is doped the exciton $iX^0$ acquires an extra (weakly bound) electron and becomes an optically active three-particle state $iX^0_e$, giving an extra peak in the absorption spectra. Notice, that the doping leads to the appearance of the  peak $X^{-}$ as well. However,  trion states $X^{-}$ with two tightly bound electrons do not have the counterpart exciton states in the undoped ML. 

The intensity of the $iX^0_e$ peak increases very slowly with $E_F$,  slower than that of the $X^{-}$ peak [Fig.~\ref{fig:spectra}c]. It is practically not visible at $E_F\lesssim 7$ meV. It should also be noted that the peak visibility declines when other relaxation mechanisms further widen the trion spectral lines.  Nevertheless, our calculations demonstrate that doping tends to enhance both $X^{-}$ and $iX^0_e$ peaks, making them dominant at sufficiently large doping levels. We also note that, remarkably, the sum of the OS's of all three peaks is practically constant being equal to that of the two-particle exciton state $X^0$, obtained from the BSE calculations and shown in Fig.~\ref{fig:spectra}d for comparison.

The spectra in Fig.~\ref{fig:spectra}a reveal  notable doping dependencies of the energy-position of the spectral peaks. Details of these dependencies are given by Fig.~\ref{fig:spectra}d. Remarkably, the doping affects the states $X^{-}$, $iX^0_e$ and $X^0_e$ in a qualitatively different way.  When the doping increases, the $X^{-}$ peak shifts to the lower energies, whereas the energy of the $X^0_e$ and $iX^0_e$ peaks increases. For comparison,  Fig.~\ref{fig:spectra}d shows the transition energies of the $X^0$ and $iX^0$ exciton states calculated for the undoped ML using the BSE. One sees that transition energies of $X^0_e$ and  $iX^0_e$ three-particle states converge to the BSE results in the limit of vanishing doping, which is another evidence of the relation between $X^0$ and $X^0_e$, as well as between $iX^0$ and $iX^0_e$ states, in this limit. Results for the $X^{-}$ and $X^0_e$ peaks positions demonstrate a perfect quantitative agreement with the available experimental data~\cite{Mak2012}, also shown in Fig.~\ref{fig:spectra}d for comparison. The data for the $iX^0_e$ peak are missing, which is probably explained by its lower OS, especially at small doping densities and large inhomogeneous broadening. 

To further understand the dependence of the three-particle energy on the doping, in Fig.~\ref{fig:separation}a we plot the energy difference $\Delta E$ between pairs of states $X^0_e - X^{-}$ and $X^0_e - iX^0_e$ as a function of the Fermi level $E_F$. The experimentally measured energy difference for states $X^0_e$ and $X^{-}$ is also plotted for comparison, showing a very good quantitative agreement with the calculations.  In all cases $\Delta E$ is a monotonically increasing function of $E_F$. 

The dashed lines in  Fig.~\ref{fig:separation}a illustrate a known simple estimate ~\cite{Burstein1954}, which yields the linear dependence $\Delta E = \Delta_0 + \alpha E_F$, with $\alpha =1$. This estimate can be rationalized by using the following intuitive arguments. As the doping increases, single particle electronic states below $E_F$ cannot contribute to the $X^0_e$ three-particle state due to the Pauli principle, thereby increasing the kinetic energy contribution to its total energy by $E_F$. On the other hand, in the $iX^0_e$ state doped electrons fill the opposite $K$ valley, as shown in Fig.~\ref{fig:band_structure}b. Therefore, the Pauli blocking does not modify the kinetic energy of the bound electron in the $-K$ valley. At the same time, for the trion state $X^-$, where both electrons are tightly bound, changes in the Fermi level does not contribute to the kinetic energy in the first order. 

Deviations from the linear estimates, seen in Fig.~\ref{fig:separation}a, are due to the potential energy contribution to the energy. In order to illustrate this, in Fig.~\ref{fig:separation}b we plot the corresponding differences between the potential energy of the same states. 
For the states $X^0_e$ and $iX^0_e$ this difference has a weaker Fermi energy dependence than that for the $X^0_e$ and $X^{-}$ pair. That explains why for  the states $X^0_e$ and $iX^0_e$ one has a better agreement with the linear dependence of the energy splitting, seen in Fig.~\ref{fig:separation}a.

\subsection{Influence of the substrate}
\label{sec:substrate}

Modifications of the substrate material give rise to changing the effective dielectric constant $\varepsilon$ which, following Eqs.~(\ref{eq:Coulomb}) and (\ref{eq:Coulomb_screened}), modifies the  effective Coulomb interaction and thus, the energy of three-particle states and the absorption spectrum. Together with the varying doping this can also be used to manipulate (engineer) optical properties of the MoS$_2$ ML structures. 

Figure~\ref{fig:epsilon}a shows binding energies of the three-particle states (relative to the $X$ exciton peak position in the undoped system) as a function of the substrate dielectric constant $\varepsilon$, calculated for two values of $E_F$. For $E_F=4.87$ meV the binding energy is a monotonically decreasing function of $\varepsilon$. However, at larger doping, $E_F=10.89$ meV,  the energy tends to saturate at $\varepsilon \gtrsim 10$. This is explained by the fact that free carriers dominate the screening of the Coulomb interaction at large $\varepsilon$, so that the role of the environmental screening  diminishes.  

In this context it is worth noting that the popular variational approach~\cite{Berkelbach2013} does not yield the saturation of the biding energy at large $\varepsilon$, because it does not take into account the screening induced by the free carriers. This explains why the deviation between the exact and the variational approach in Fig.~\ref{fig:epsilon} is relatively small for the undoped system, but grows at larger doping thereby limiting applicability of this approach for heavily doped TMDC ML's.

The dependence of the peak intensity on the dielectric constant, shown in Fig.~\ref{fig:epsilon}b and c, demonstrates qualitatively different behaviour for different three-particle states.  While the intensity of the $X^0_e$ peak is a monotonically decreasing function of $\varepsilon$, the intensities of the $iX^0_e$ and $X^-$ peaks increase at small but saturate at large values of $\varepsilon$. 

The decreasing intensity of the $X^0_e$ peak is attributed to the increasing radius of the 
three-particle state which results in a smaller spatial overlap between the localized electron and the localized hole.  In the case of the $iX^0_e$ state, the optical transition is determined by the overlap between the wave functions of the delocalized electron and the localized hole. As the localization length of the hole increases its overlap with the delocalized electron increases as well, leading to the larger OS. We note in passing, that the total OS from all three peaks coincides with that of the $X^0$ exciton state in the undoped system obtained by solving the BSE (not shown here).

\section{Conclusions}
\label{sec:conclusions}

This work presented results of the theoretical investigation of the linear absorption spectrum of a doped MoS$_2$ ML. The calculations are done using a combination of the {\it ab initio} approach for the band structure and the solution of the three-particle problem, where the doping is accounted for by discretizing the phase-space. This method allowed us to study the doping dependence of the lowest spectral peaks associated with three qualitatively different three-particle states $X^-$, $X^0_e$, and $iX^0_e$. The approach is general an can be extended to other TMDC ML's. 

The calculations reproduce available experimental data for two spectral peaks $X^-$ and $X^0_e$ with an excellent accuracy. We also predict that at sufficiently large doping the spectrum acquires one more peak due to the intervalley exciton state $iX^0_e$, which is dark in the undoped ML but is brightened when the doping is large. Observations of the predicted here $iX^0_e$ state is related to the sample quality, in which the peak widths should be smaller than the energy difference between the $X^0_e$ and $iX^0_e$ peaks shown in Fig.~\ref{fig:separation}a. However, an indirect evidence such as OS doping dependencies of the spectrally resolved peaks may serve as an experimental signatures of the $iX^0_e$ states. It should be noted that transitions between other multi-particles states can, in principle, contribute to the spectral interval between the $X^0_e$ and $X^-$ peaks, for example, transitions between exciton $X^0$ and bi-exciton $X^0X^0$ states.~\cite{Hao2017,Steinhoff2018,Zhang2015} However, in the light absorption processes such transitions yield only non-linear contributions and are missing in the linear absorption spectrum, calculated here.

We also demonstrate how contributions by different excitonic states to the spectrum can be manipulated by changing the dielectric environment.  Our qualitative conclusions are general and should hold for other TMDC layered structures with the valley degeneracy in the single-particle spectrum.

\section*{Acknowledgements}
The {\it ab initio} calculations were supported by the Russian Science Foundation under the grant 18-12-00429. BSE calculations were supported by the Deutsche Forschungsgemeinschaft (DFG, German Research Foundation – Project-ID 314695032 – SFB 1277). A.V. thanks Prof. Cid Araujo for inspiring discussions. 


\bibliography{bibliography}

\end{document}